# Foundational Techniques for Wireless Communications: Channel Coding, Modulation, and Equalization

Solomon McKiernan
*Unaffiliated*
Bedford, UK
0009-0000-2352-3769

*Abstract*—This paper analyses foundational techniques for improving wireless communication systems, including coding methods, modulation schemes, and channel equalization. Using industry-standard simulation tools, the paper evaluates the performance of these techniques under different channel conditions. Convolutional codes, punctured and unpunctured, are assessed for reliable data transfer. The suitability of various modulation schemes, such as Phase Shift Keying (PSK) and Quadrature Amplitude Modulation (QAM), are examined. Linear and decision-feedback equalization techniques are evaluated for mitigating the effects of channel impairments. The paper provides practical insights into the implementation of these techniques, emphasizing their importance in modern wireless communication systems.

## I. INTRODUCTION

Reliable data transfer is a crucial requirement for modern wireless communication systems. In this paper, an analysis of foundational techniques used in wireless communication systems to achieve reliable data transfer is presented. The focus of this analysis is on the use of coding methods and industry-standard simulation to evaluate the performance of modulation schemes and channel equalization techniques. These techniques have been assessed against standard key performance indicators for wireless communication systems, namely the signal-to-noise ratio (SNR) and bit error rate (BER).

This paper is grounded in the theoretical framework of information theory and signal processing, offering practical implications into utilizing these techniques in modern wireless communication systems. By analysing and evaluating these techniques, the significance of deliberate selection and assessment of coding, modulation, and channel equalization techniques in designing wireless communication systems is highlighted.

## II. CONVOLUTIONAL CODING OVERVIEW

Convolutional encoding is widely used in wireless communication systems to ensure reliable data transfer. It employs a shift register to perform bitwise operations on the input data stream, generating an encoded output to be transmitted based on the contents of the registers and feedback connections. The received signal is then typically decoded using a Viterbi decoder to recover the original data.

By operating on a sliding window of input data bits, convolutional encoding outputs a set of encoded bits that are wider than the original input due to the added error correction. This wider bit stream adds redundancy to the data and improves the robustness of the transmitted signal, in turn this reduces the bit error rate (BER).

This paper presents a detailed analysis of convolutional encoding, including its mathematical foundations, practical implementation, and application in modern wireless communication systems. It demonstrates how convolutional encoding can improve the reliability of data transmission by reducing the effect of noise and interference in the communication channel. To achieve this, one of the primary methods used is simulation in MATLAB, an industry-standard software package, in conjunction with their Communications Toolbox.

This paper also explores the use of the punctured coding system, which allows for the encoding and decoding of higher rate codes using standard rate 1/2 encoders and decoders. Punctured convolutional encoding involves selectively removing certain bits from the resulting code after performing convolutional encoding on the original code. Further explanation and demonstration of this process is shown in the following sections.

## III. CHANNEL CODING MODELS

Puncturing is used in convolutional coding to selectively remove some of the parity bits from the code to adjust the code rate to better suit the channel. The use of transpose matrix is important for punctured convolutional coding because it allows for the systematic encoding of the code and helps reduce the complexity of the decoder. The transpose matrix is used to puncture the code by removing specific rows, and to recover the original data by reversing the puncturing process during decoding.

The Punctured Convolutional Coding model in Fig. 1 demonstrates a punctured coding system that uses rate 1/2 convolutional encoding and Viterbi decoding. The model simulates the transmission of a convolutionally encoded binary phase shift keying (BPSK) signal through an additive white Gaussian noise (AWGN) channel, and then recovers the original uncoded signal using Viterbi decoding.

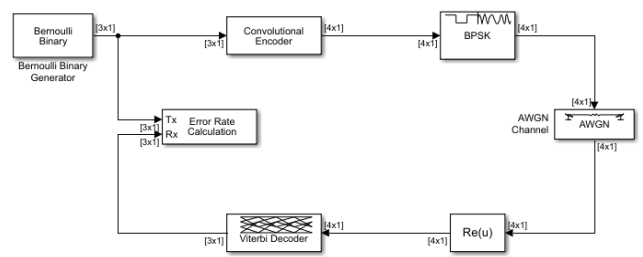

Fig. 1. Punctured Convolutional Coding Model in Simulink

The model computes the error rate by comparing the original signal and the decoded signal. This demonstrates how the puncturing technique can enable encoding and decoding of higher rate codes using standard lower rate coders and provides a practical example of how this technique can be implemented in modern wireless communication systems using the Simulink software package.

A custom script was written to automate a sweep of $E_b/N_0$ (dB) values, generating the dataset plotted on Fig. 2. When the $E_b/N_0$ is at a value of 6 dB or greater, practically no bit errors are observed. Hence, this shows with a suitably a high SNR, and therefore transmit power, one can effectively eliminate AWGN-induced errors. This trend of increasing $E_b/N_0$ with decreasing BER is demonstrated throughout this paper. While this general trend is well established, the relative power required of different techniques is of practical significance. Under the majority of conditions shown, forward error correction (FEC) reduces the power required for a given BER.

To confirm the validity of these results, they were compared with an established performance bound as defined in [1]. The BER performance of a rate $r = (n-1)/n$ punctured code is bounded by (1).

$$P_b \leq \frac{1}{2(n-1)} \sum_{d=d_{free}}^{\infty} \omega_d \, erfc(\sqrt{rd(E_b/N_0)}) \qquad (1)$$

The expression involves several parameters and is used to determine the theoretical maximum performance of a punctured code with the aforementioned rate. It consists of a sum of terms that depend on a parameter called $\omega_d$ where $d$ d is the degree of the polynomial used to generate the punctured code. Additionally, the expression includes the complementary error function ($erfc$) of the square root of a parameter that takes into account the received signal energy per bit (Eb) and the noise spectral density (N0). For more information, see [1]. Using a larger $E_b/N_0$ to gain a better overview, the BER vs $E_b/N_0$ plot shown in Fig. 2 was generated, with validity proven by the bound. Fig. 2 displays the relationship between the BER and $E_b/N_0$ of a punctured code. As $E_b/N_0$ increases, BER decreases, demonstrating an inverse relationship between the two variables. The steepness of the roll-off is important in that it reduces the Eb/No at which you can achieve an arbitrarily low BER.

There are two main types of channel coding: convolutional coding, as aforementioned, and block coding. Block coding involves dividing the data into blocks of fixed size and adding parity check bits to each block. These parity check bits are calculated from the data in the block and are used to detect and correct errors. The main advantage of block coding is its simplicity and ease of implementation. Reed-Solomon (RS) codes are a commonly used type of block code. RS codes typically require larger word sizes compared to convolutional codes so are more suitable for applications with higher data rates.

While both types of codes can provide error-correction capabilities, they have different properties that make them more suitable for certain applications. Block codes are typically used in applications that require high levels of error correction, such as satellite and deep space communication. This is due to block codes working by dividing the message into fixed-length blocks and adding redundancy to each block with their most important feature being the size of the blocks can be readily increased to obtain better performance. Convolutional codes, on the other hand, are better suited for applications where minimal implementation costs are a priority, such as WLANs with lower error-correction capabilities. This is due to convolutional codes working by encoding the message as a continuous stream of bits, using a shift register and a feedback function. Simulations were run over the same $E_b/N_0$ range, keeping the code rate of 5/7, QPSK modulation scheme, and data input seed constant; see Fig. 3.

Fig. 3 shows a simulation of RS outperforming convolutional codes at low to medium $E_b/N_0$, while convolutional codes outperform RS codes at higher $E_b/N_0$. This trend occurs because RS codes can correct a fixed number of errors per block, while convolutional codes can correct errors continuously. As the SNR decreases, the probability of having a larger number of errors in the block increases, and the RS code becomes less effective. At high SNR, the error rate is low, and convolutional codes, with their continuous error correction capability, can take advantage of this to increasingly achieve better performance. It is important to note that generally only BERs under $10^{-3}$ are considered acceptable for reliable transmission, in which case convolutional is shown to outperforms RS in this instance.

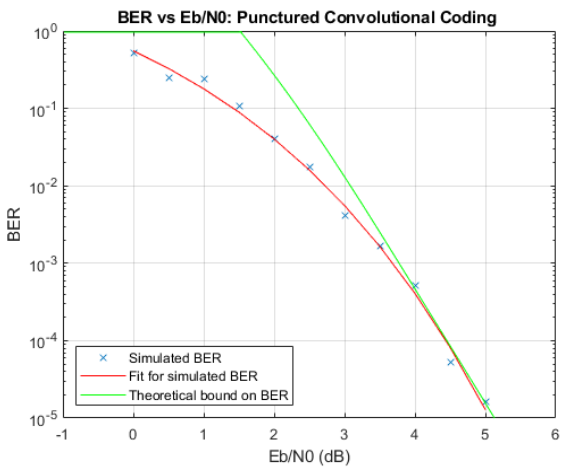

Fig. 2.  BER vs $E_b/N_0$: Punctured Convolutional Coding

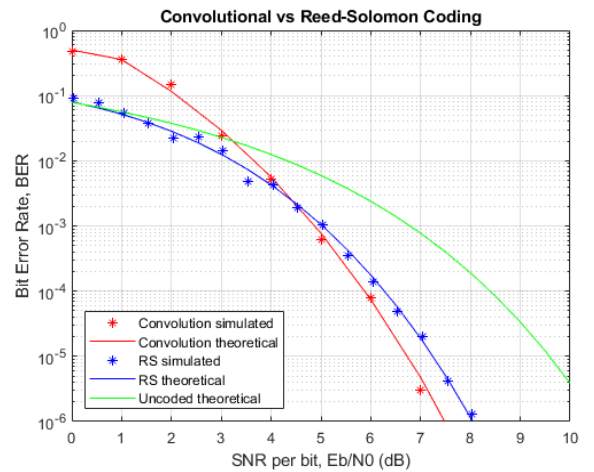

Fig. 3.  Convolutional vs Reed-Solomon Coding BER plot

The plot also shows both coded systems performing worse than uncoded systems at very low SNR. This is most likely due to the coding overhead reducing the signal power, making it more difficult to detect the signal at the receiver. As the SNR increases, the benefits of coding progressively outweigh the overhead, and the coded system begins to outperform the uncoded system as intended.

While an RS block size was selected to provide a fair baseline comparison, with higher block sizes, RS code would be expected to increasingly outperform convolutional code.

## IV. Signal to Noise ratio of Radio LAN

To analyse a communication link with increased complexity, and therefore more akin to real-world modern applications, a Simulink model was created based on the High-Performance Radio Local Area Network (HIPERLAN/2) which is described in the European Telecommunications Standards Institute's (ETSI) specification for high-rate wireless Local Area Networks (WLAN). Despite HIPERLAN/2 itself being outdated, it shares many key fundamental blocks with modern wireless communication systems, while still allowing for clear analysis of modulation, coding, and channel equalization techniques. These techniques, such as orthogonal frequency division multiplexing (OFDM), are still utilised in contemporary radio technologies such as IEEE 802.11ax [2], marketed as WI-FI 6E.

The model employs OFDM in the 5 GHz band, offering raw data rates up to 54 Mbps. Using this model, the transmitter-side channel coding and modulation for the 16-QAM, ¾ code rate mode, along with an ideal receiver chain and AWGN channel was demonstrated. The model was based on the HIPERLAN/2 simulation setup shown in [3], updating various blocks and replacing fixed values with workspace variables allowing automated sweeps of the model to generate comparison plots. Simulation model discussed is shown in Fig. 4. The link with an SNR set to 25 dB produced the spectrum plot shown in Fig. 5 along with the Constellation Diagram in Fig. 6.

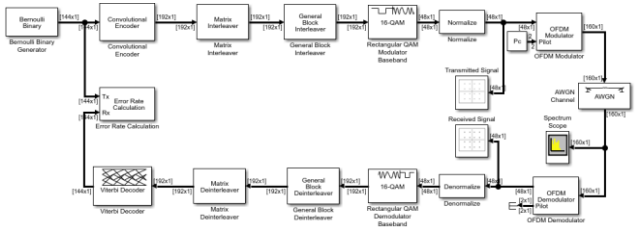

Fig. 4. HIPERLAN/2 based Simulink model

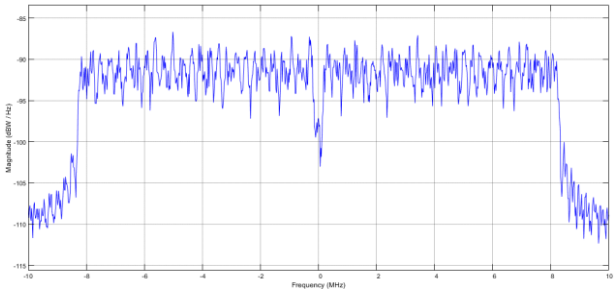

Fig. 5. HIPERLAN/2 Spectrum Plot

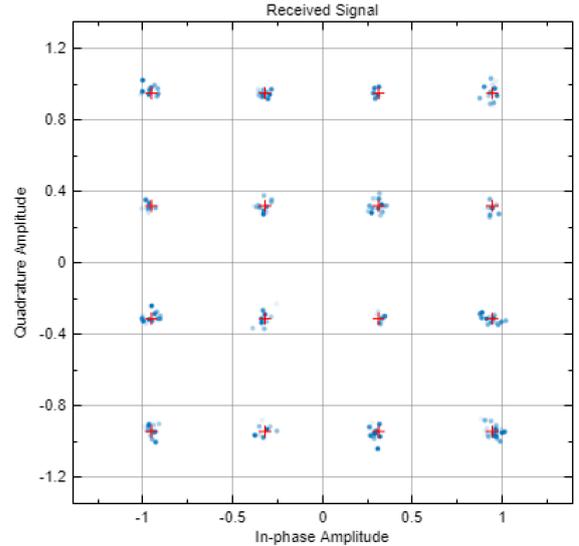

Fig. 6. HIPERLAN/2 Constellation Plot

Through simulations of this model, one can gain a better understanding of the performance of wireless LANs under different coding and modulation schemes, which is useful for future wireless communication system designs.

The Spectrum Scope showed a signal with a real, i.e., above 0Hz, bandwidth of 8 MHz. In this range the signal power spectral density magnitude fluctuates around -90 dBW/Hz, approximately 25 dB above the noise floor as specified. The frequency spectrum plot for the OFDM signal is not flat due to the channel induced noise.

The Constellation Diagram, Fig. 6, shows that the channel noise has affected the signal quality, causing errors in the received signal with less sharp points i.e., larger clusters.

The Error Vector Magnitude (EVM) is a measure of how much a received constellation point deviates from the ideal reference value for a given modulation index & can be determined from the Constellation Diagram using (2). The maximum permissible EVM is set by the IEEE and for this system, 16-QAM modulation with ¾ code rate, is -19dB [4].

$$EVM = \frac{\sqrt{\sum_{i=1}^{L_p}\sum_{j=1}^{N_c}|R_{i,j}-S_{i,j}|^2}}{\sqrt{\sum_{i=1}^{L_p}\sum_{j=1}^{N_c}|S_{i,j}|^2}} \qquad (2)$$

This calculation considers the number of frames ($L_p$), number of carriers ($N_c$), received signal ($R_{i,j}$) and ideal symbol location ($S_{i,j}$). For evaluating a single received point, the number of frames and carriers are set to one, resulting in (3) which has been expressed in logarithmic units to allow for direct evaluation against the upper permissible limit.

$$EVM_{pt.(dB)} = \frac{|R_{i,j}-S_{i,j}|}{|S_{i,j}|} \qquad (3)$$

Using the point furthest from its respective reference shown in the top-left decision region of Fig. 6, a value of $-29dB$ can be calculated, far below the -19dB limit. This demonstrates that the system in Fig. 4 has high signal quality due to the use of channel coding and modulation.

Running a sweep of AWGN SNR values -5 to 25 dB with a constant channel input signal power 0.01 W generated the values shown in Table 1.

TABLE I.    HIPERLAN/2 SIMULATED SNR AND BER VALUES

| SNR (dB) | Calculated BER | Number of Errors | Number of bits |
|----------|----------------|------------------|----------------|
| -5 | 0.5014 | 180551 | 360110 |
| 0 | 0.5001 | 180127 | 360110 |
| 5 | 0.4975 | 179151 | 360110 |
| 10 | 0.4876 | 175603 | 360110 |
| 15 | 0.1904 | 68549 | 360110 |
| 20 | 0 | 25 | 360110 |
| 25 | 0 | 0 | 360110 |

Plotting this data with, including intermittent datapoints, gave Fig. 7. Note the distinction between SNR and $E_b/N_0$ becomes more relevant when comparing various modulation indexes, i.e., differing spectral efficiencies, $n$, due to (4).

$$SNR = \left(\frac{E_b}{N_0}\right) \cdot n \tag{4}$$

Spectral, also referred to as bandwidth, efficiency is the information rate that can be transmitted over a given bandwidth in a system. In general, the higher the modulation order, M, the higher the spectral efficiency. Higher-order modulations allow for more bits to be transmitted per symbol, and therefore more data to be transmitted in a given bandwidth. However, as the modulation order increases, the required SNR or $E_b/N_0$ for reliable transmission also increases, which can limit the achievable spectral efficiency.

As anticipated, the correlation between Signal-to-Noise Ratio (SNR) and Bit Error Rate (BER) in the 16-QAM graph follows an inverse proportion. Put simply, as the SNR increases, the BER decreases and vice versa, the same trend as shown in Fig. 2. This is because a higher SNR indicates a more robust signal relative to noise, leading to fewer transmission errors. Conversely, a lower SNR denotes a weaker signal compared to noise, making it harder to differentiate between the signal and noise and elevating the probability of errors.

Consequently, the SNR is a critical factor influencing communication quality in a 16-QAM system, and a suitably high SNR is essential to minimize BER and guarantee dependable data transmission.

Simulations were also run to explore usage of different convolutional coding schemes, these used two standard code rates $R = 1/2$ and $R = 3/4$; defined in [4].

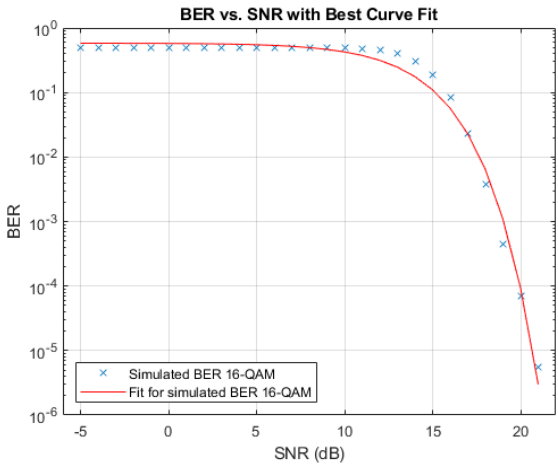

Fig. 7.   HIPERLAN/2 BER vs SNR simulated plot

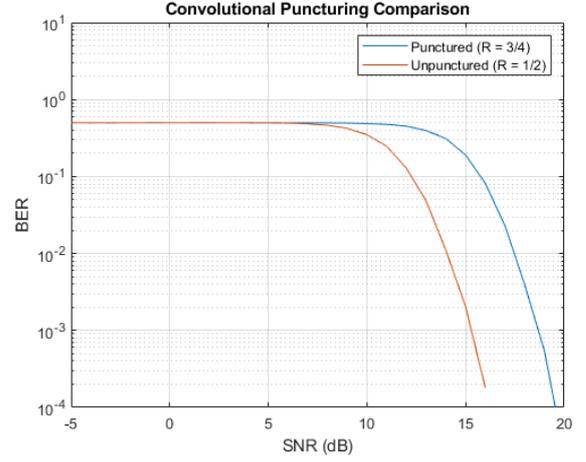

Fig. 8.   Convolutional punctured vs unpunctured

This specific pair was chosen because one could be achieved with puncturing and the other without. The resulting plot is shown in Fig. 8. This plot demonstrates that while puncturing code, increasing the Code Rate, can transmit more data, it requires more signal power to achieve the same BER.

The unpunctured code still has Convolutional Encoding resulting in a Code Rate of 1/2 i.e., 50% of the transmitted bits are used for error correction, making it still more resilient to noise than default. The punctured code allows more data to be transmitted, using a Punctuation Vector of is [1 1 1 0 0 1]′ which has a Puncturing Rate of 4/6, resulting in an overall Code Rate of 3/4 i.e., only 25% redundancy bits.

The Convolutional Encoder block is a key component in the model, which enhances the error correction capabilities of the system. This block processes a binary input sequence and generates a binary output. The block applies a convolutional code to the input sequence, which adds redundancy to the data to make it more resilient to errors.

To configure the Convolutional Encoder block, the puncture vector and the Trellis structure parameters must be set. The puncture vector is a pattern of 1s and 0s that indicates the kept and punctured bits. In the model, the puncture vector is set to [1 1 1 0 0 1]′, where the 0s are the punctured bits. The Trellis structure parameter specifies the encoder using its constraint length, generator polynomials, and possibly feedback connection polynomials.

The Trellis structure is set to $poly2trellis(7, [133\ 171])$ which means that it uses an encoder with a constraint length of 7 and code generator polynomials of 171 and 133 in octal notation.Constraint length determines the number of previous input bits that affect the current output bit while the generator polynomials are used to define the mapping between the input and output of the convolutional encoder.

## V.   MODULATION SCHEMES

Modulation schemes are used to encode data onto a carrier wave before transmission. The choice of modulation scheme affects the bandwidth efficiency, robustness to noise, and data rate of the system; this is elaborated on in [5]. As mentioned prior, a key measure of the performance of a modulation scheme is its BER, which is the probability of an error in a single bit transmission. Different modulation schemes have different BER characteristics, and it is important to compare them to determine their suitability for a given application.

Therefore, several modulation schemes commonly used in wireless communication systems were simulated with the results presented in a BER comparison. To achieve this, the modulation and demodulation block were replaced within the model shown in Fig. 4 along with iterative adjustments to the modulation index defined in the code. All blocks requiring parameter values derived from the modulation index, such as the OFDM modulator, were updated with the change in the global $M$ variable correspondingly. The resultant BER plot is shown in Fig. 9.

Fig. 9 shows at lower values of $E_b/N_0$, the response is relatively flat due to the noise floor dominating the received signal resulting in errors occurring for even low levels of modulation. As the $E_b/N_0$ increases, the SNR also increases, and the probability of bit errors decreases, resulting in a lower BER. This occurs until a certain threshold, known as the "knee point," is reached, which typically occurs around +5dB to +10dB for many communication systems, corresponding with what was simulated. Beyond this point, the slope of the BER vs $E_b/N_0$ curve becomes steeper, indicating a faster decrease in the probability of bit errors as the SNR increases.

Fig. 9 provides important insights into the Bit Error Rate (BER) performance of different Modulation Indexes (M) of 2, 4, 8, 16, 64, and 256, at varying $E_b/N_0$ (dB) values. As M increases, BER performance worsens due to higher sensitivity to noise and interference.

The lowest modulation schemes such as BPSK achieve lower BER with the least $E_b/N_0$, whereas the highest modulation scheme, 256-QAM, requires the largest $E_b/N_0$ to achieve a given BER. However, typically the higher M the more information can be transmitted given a suitable $E_b/N_0$ therefore resulting in increased spectral efficiency. By comparing these schemes, one can determine the most suitable option for a given application. System requirements for BER and spectral efficiency along with available transmit power, relating to $E_b/N_0$, are key deciding factors when selecting a modulation scheme. In summary, Fig. 9 highlighting the trade-off between performance and spectral efficiency in M-PSK and M-QAM modulation. Designers should carefully select the modulation scheme that best meets their needs while balancing these factors. For example, if large amounts of data are required and there is substantial transmit power available, to achieve a desired BER, then 256-QAM may be a suitable choice.

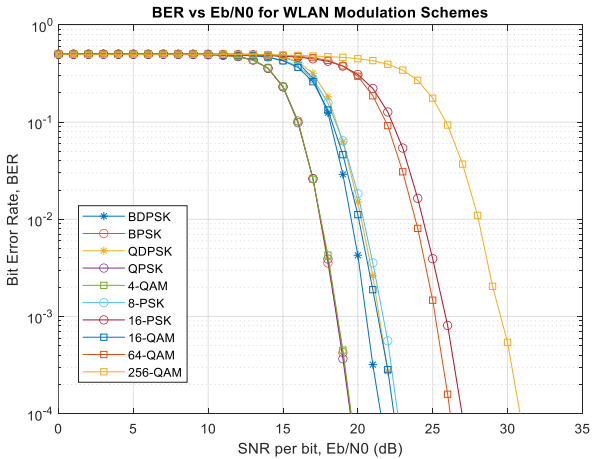

Fig. 9. HIPERLAN/2 based BER plot for various modulation schemes

IEEE 802.11ax (WI-FI 6E) only includes BPSK, QPSK, 16-QAM, 64QAM, 256QAM [4] but additional schemes have been included for comparison and to highlight why their respective trade-offs between data capacity and signal power have resulted in them being excluded from the standard. Wi-Fi 6E introduced 256-QAM, which offers higher data rates at the expense of increased susceptibility to noise and interference, a trend clear in Fig. 9.

To generate the data plotted in Fig. 9 the OFDM Fast Fourier Transform (FFT) length was changed from 128 to 256, both standard values from [4] chosen to simplify the FFT, in order to accommodate the BPSK scheme. This was required as the total number of OFDM carriers, $N_{SC}$, is in fact equal to the FFT length. However, not all of these subcarriers can be used for data transmission, $N_U$, as some are reserved for other purposes, such as subcarrier pilots, $N_P$, and DC null subcarrier, $N_{DC}$; the DC null in particular can be seen in Fig. 5. For the subsequent simulations, only two guards were used, with the first guard calculated as shown in (5) and the second guard simply set as one less than this.

$$G = \frac{N_{SG} - N_U - N_P}{2} \tag{5}$$

To complete a fair comparison, the normalization factor for the PSK schemes was set to a constant value; this is due to M-PSK having a constant amplitude, and therefore constant average power per symbol, just varying phases. Conversely, the QAM schemes had a varying normalization factor dependent on M, shown in, due to M-QAM having multiple amplitude levels and therefore average powers. These normalization factors ensured that the transmit power was constant regardless of the modulation scheme.

## VI. EQUALIZATION

Equalizers are signal processing blocks that are used to compensate for channel distortions in wireless communication systems. The distortions arise due to multipath propagation, frequency-selective fading, and other impairments in the wireless channel. Essentially, they estimate the channels frequency response in order to apply the inverse, cancelling out the distortions. In WLAN systems, equalizers can help improve the performance of the receiver by mitigating the effects of these distortions leading to improved BER. However, they are not always necessary, and their use depends on the specific communication standards and system requirements. For example, HIPERLAN/2 and IEEE 802.11ax [4] use techniques like OFDM and MIMO to overcome channel impairments, which obviate the need for equalizers.

However, WLANs may still require an equalizer in scenarios where there is a significant amount of channel distortion or interference. For example, in indoor environments with obstacles such as walls or in outdoor environments with multiple reflective surfaces, the transmitted signal can experience multipath fading, where the signal arrives at the receiver via multiple paths with different delays and amplitudes. This can cause inter-symbol interference (ISI), where the symbols transmitted in one time interval interfere with the symbols transmitted in the adjacent time intervals.

In such scenarios, an equalizer can be used to mitigate the effect of ISI and improve the bit error rate (BER) performance. However, in modern systems, OFDM is commonly selected as the primary technique used to combat ISI. While OFDM can practically eliminate ISI, when the maximum multipath delay is less than the guard interval, equalizers can still be useful to mitigate frequency selective fading.

It is important to compare equalizer types for WLAN as different equalizers have different complexity, performance, and adaptability trade-offs. For example, decision-feedback equalizers (DFE) can provide excellent performance but at the cost of higher complexity compared to linear equalizers. Similarly, adaptive equalizers can better handle time-varying channels but require additional training overhead. Understanding these trade-offs and choosing the right equalizer type for a particular WLAN scenario can lead to better performance and spectral efficiency. Modifying the code provided in [6], a maximum likelihood sequence estimation (MLSE) equalizer is demonstrated estimating a channel frequency response, see Fig. 10. Increasing the range of [6] resulted in Fig. 11.

Fig. 10 demonstrates that these equalizers can dynamically provide a suitably accurate estimation of the channel which in turn allows them to attempt to compensate for any distortions. This was run at an $E_b/N_0$ of 14dB and resulted in a BER well below $10^{-3}$ shown in Fig. 11. As the $E_b/N_0$ increases, the channel estimation of MLSE generally becomes more accurate. This is because at higher $E_b/N_0$, there is less noise and the received signal is more reliable. As a result, the channel can be estimated with greater accuracy, leading to improved performance of the MLSE equalizer. Fig. 11 presents a comparison of three different equalizer types based on their BER performance under a range of $E_b/N_0$ conditions.

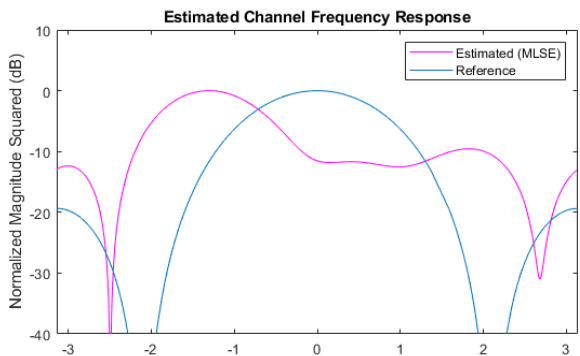

Fig. 10. MLSE Equalizer Estimated Channel Response

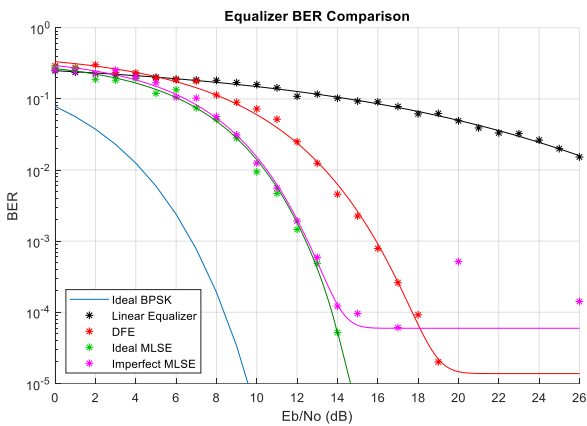

Fig. 11. Equalizer BER comparison through AWGN using BPSK

It is observed that the MLSE equalizer performs the best across all SNR values, while the linear equalizer has the worst performance. The DFE falls in between these two but is slightly closer to the MLSE equalizer. The performance of the linear equalizer tends to diverge with increasing $E_b/N_0$, which suggests that it may be better suited for low-power systems particularly where complexity is a concern.

For high-power systems or scenarios with more severe channel distortions, more advanced equalizers such as the DFE or MLSE equalizer may be necessary to achieve acceptable performance levels. On the other hand, the DFE strikes a good balance between performance and complexity. While it requires more computational resources than the linear equalizer, it is still more practical than the MLSE, known for its high complexity and computational requirements.

Selecting the right equalizer for a WLAN system requires careful consideration of multiple factors, including the specific channel characteristics, the desired data rates, power consumption, and computational resources. Fig. 11 provides an indication of how each type of equalizer will perform comparatively and highlights the trade-off between performance and complexity.

## VII. Summary

The contents of this paper demonstrates that the selection and design of coding methods, modulation schemes, and equalizers are critical for achieving the desired data rates and quality of service in modern wireless communication systems and highlights some of the most important trade-offs between popular techniques. These techniques must be chosen based on the specific requirements of the wireless communication system, such as power consumption, computational resources, and channel characteristics. Through simulations using industry-standard system modelling software, this paper has demonstrated the effectiveness of various coding and modulation techniques, as well as the suitability of different equalization techniques for wireless communication systems.

## VIII. References

[1] Y. Yasuda, K. Kashiki and Y. Hirata, "High Rate Punctured Convolutional Codes for Soft Decision Viterbi Decoding," *IEEE Transactions on Communications*, Vols. COM-32, pp. 315-319, 1984.

[2] E. Khorov, A. Kiryanov, A. Lyakhov and G. Bianchi, "A Tutorial on IEEE 802.11ax High Efficiency WLANs," *IEEE Communications Surveys & Tutorials*, vol. 21, pp. 3-4, 2019.

[3] K. Bondre and S. Rathkanthiwar, "Performance of HIPERLAN/2 for 16QAM, with ¾ code rate with and without Rapp's model," *International Journal of Computer and Communication Technology*, vol. 2, no. 2, pp. 3-4, 2011.

[4] IEEE, "IEEE Standard for Information Technology--Telecommunications and Information Exchange between Systems - Local and Metropolitan Area Networks--Specific Requirements - Part 11: Wireless LAN Medium Access Control (MAC) and Physical Layer (PHY) Specifications," *IEEE Std 802.11-2020*, pp. 1-4379, 2021.

[5] J. Proakis and M. Salehi, Fundamentals of Communication Systems 2nd Ed., NJ, USA: Pearson Education, 2014.

[6] Mathworks, "BER Performance of Different Equalizers," [Online]. Available: https://uk.mathworks.com/help/comm/ug/ber-performance-of-different-equalizers.html. [Accessed 20 March 2023].